\documentclass[prb,aps,showpacs,superscriptaddress,twocolumn]{revtex4}
\usepackage{latexsym}
\usepackage{amsfonts}
\usepackage{amssymb}
\usepackage{amsmath}
\usepackage{graphicx}
\begin{document}
\title{Collective modes and quasiparticle interference
on the local density of states of cuprate superconductors}
\date{\today}
\author{C.-T. Chen}
\author{N.-C. Yeh}

\affiliation{Department of Physics, California Institute of
Technology, Pasadena, CA 91125, USA}

\begin{abstract}
The energy, momentum and temperature dependence of the
quasiparticle local density of states (LDOS) of a two-dimensional
$d_{x^2-y^2}$-wave superconductor with random disorder is
investigated using the first-order T-matrix approximation. The
results suggest that collective modes such as spin/charge density
waves are relevant low-energy excitations of the cuprates that
contribute to the observed LDOS modulations in recent scanning
tunneling microscopy studies of $\rm Bi_2Sr_2CaCu_2O_x$.

\end{abstract}
 \maketitle

One of the most widely debated issues in cuprate superconductivity
is the possibility of preformed Cooper pairs~\cite{Anderson87,Dagotto92,Emery97,ChenQJ99}
and the origin of the pseudogap phenomenon.~\cite{Timusk99,Yeh02a}
Recent experiments have demonstrated that the pseudogap
phenomenon is unique to the hole-doped
(p-type) cuprates and is absent above $T_c$ in electron-doped
(n-type) cuprates.~\cite{Chen02,Kleefisch01,Yeh03a} Furthermore,
in the quasiparticle tunneling spectra of the double-layer
$\rm Bi_2Sr_2CaCu_2O_{8+\delta}$ (Bi-2212)~\cite{Krasnov00} and the one-layer
$\rm Bi_2(Sr_{2-x}La_x)Cu_2O_{6+\delta}$ (Bi-2201)~\cite{Yurgens03}
systems, it is shown that the pseudogap can be distinguished from
the superconducting gap: the former evolves smoothly with
increasing temperature whereas the latter vanishes at $T_c$.
These phenomena suggest that the pseudogap
may be associated with a competing order~\cite{Yeh02a,Yeh03b}
that coexists with the superconducting phase for $T < T_c$ and persists above
$T_c$ until a pseudogap temperature $T^{\ast}$. The competing quantum ordered
phase~\cite{Sachdev03,Kivelson03} can be manifested in the form of collective modes
such as charge- and spin-density waves (CDW and SDW) in the superconducting
state, as inferred from neutron scattering experiments in a variety of p-type
cuprates.~\cite{Kastner98,Fong99a,Mook02,Fujita02,Lake01}
However, whether these collective modes are closely correlated with
superconductivity remain controversial. Recent scanning tunneling spectroscopic
studies of the Fourier transformed (FT) quasiparticle local density of states
(LDOS) of Bi-2212~\cite{Hoffman02a,McElroy03,Howald03}
have stimulated further discussions on the relevance of collective
modes.~\cite{Polkovnikov02a,Polkovnikov03,Podolsky03,ZhangD03,WangQH03,Capriotti03}
While Bogoliubov quasiparticle interference apparently plays an important
role in the observed FT-LDOS in the superconducting state, certain spectral
details of the LDOS cannot be accounted for unless collective modes
are considered.~\cite{Polkovnikov02a,Polkovnikov03,Podolsky03} In particular,
the findings of 4 high-intensity Bragg peaks remaining above $T_c$ in the
FT-LDOS map of Bi-2212~\cite{Yazdani03} cannot be reconciled with
quasiparticles being the sole low-energy excitations. These new developments
motivate us to reexamine the role of collective modes in cuprate
superconductors by considering the energy ($E$), momentum transfer
($q$) and temperature ($T$) dependence of the resulting FT-LDOS modulations.

We begin our model construction by noting that substantial
quasiparticle gap inhomogeneities are observed in the low-temperature
tunneling spectroscopy of under- and optimally doped Bi-2212 single
crystals,~\cite{Howald01,Lang02} suggesting at least two types of spatially
separated regions, one with sharp quasiparticle coherence peaks at
smaller energies $\Delta _d$ and the other with rounded hump-like
features at larger energies $\Delta ^{\ast}$. On the other hand,
low-energy LDOS (for $E <~ 0.5 \Delta _d$) of Bi-2212 exhibit
long-range spectral homogeneity. We therefore conjecture that
dynamic SDW or CDW coexist with cuprate superconductivity and that
they are only manifested in the quasiparticle LDOS when pinned by
disorder. Thus, regions with rounded hump features in the
quasiparticle spectra are manifestation of localized charge
modulations due to pinning of collective modes by disorder, and
the wavevector of the charge modulation is twice of that for the
collinear SDW order, as proposed in
Refs.~\onlinecite{Polkovnikov02a,Polkovnikov03,Zachar98}. In contrast,
regions with sharp quasiparticle spectral peaks are representative
of generic Bogoliubov quasiparticle spectra with a well-defined
$d$-wave pairing order parameter $\Delta _\textbf{k} \approx
\Delta _d \cos 2\theta _\textbf{k}$, where $\Delta _d$ is the
maximum gap value and $\theta _\textbf{k}$ is the angle between
the quasiparticle wavevector \textbf{k} and the antinode
direction. Our model therefore assumes `puddles' of spatially confined
`pseudogap regions' with a quasiparticle scattering potential
modulated at a periodicity of 4 lattice constants along the Cu-O
bonding directions, and the spatial modulations can be of either
the `checkerboard' pattern~\cite{Polkovnikov02a,Polkovnikov03} or `charge
nematic' with short-range stripes.~\cite{Yeh03b,Kivelson03} In the
limit of weak perturbations, we employ the first-order T-matrix
approximation and consider a ($400 \times 400 $) sample area with
either 24 randomly distributed point impurities or 24 randomly
distributed puddles of charge modulations that cover approximately
6$\%$ of the sample area. For simplicity, we do not consider the
effect of disorder on either suppressing the local pairing
potential $\Delta _d (r)$ or altering the nearest-neighbor hopping
coefficient ($t$) in the band structure of Bi-2212,
although such effects reflect the internal structures of charge
modulations.~\cite{Podolsky03,ZhangD03}

Specifically, the Hamiltonian of the two-dimensional
superconductor is given by ${\cal H} = {\cal H} _{BCS} + {\cal H}
_{imp}$, where ${\cal H} _{BCS}$ denotes the unperturbed BCS
Hamiltonian of the $d$-wave superconductor, ${\cal H} _{BCS} =
\sum _{\textbf{k} \sigma} (\epsilon _k - \mu )
c^{\dagger}_{\textbf{k} \sigma} c_{\textbf{k} \sigma} + \sum
_{\textbf{k}} \Delta _\textbf{k} \lbrack c^{\dagger}_{\textbf{k}
\uparrow} c^{\dagger} _{-\textbf{k} \downarrow} + c_{-\textbf{k}
\downarrow} c_{\textbf{k} \uparrow} \rbrack$, and ${\cal H}_{imp}$
is the perturbation Hamiltonian associated with impurity-induced
quasiparticle scattering
potential.~\cite{Polkovnikov03,ZhangD03,WangQH03,CDW} Using the
T-matrix method, the Green's function ${\cal G}$ associated with
${\cal H}$ is given by ${\cal G} = {\cal G} _0 + {\cal G} _0 T
{\cal G} _0$, where ${\cal G} _0$ is the Green's function of
${\cal H} _{BCS}$ and $T = {\cal H} _{imp} / (1-{\cal G} _0 {\cal
H} _{imp})$. The Hartree perturbation potential for single
scattering events in the diagonal part of ${\cal H}$ and for
non-interacting identical point impurities at locations
$\textbf{r} _i$ is $V_{\alpha}(\textbf{q}) = \sum _i V_{s,m} e^{i
\textbf{q} \cdot \textbf{r}_i}$ for non-magnetic ($V_s$) and
magnetic ($V_m$) impurities~\cite{WangQH03}, whereas that for
puddles with short stripe-like modulations centering at
$\textbf{r}_j$ is~\cite{CDW}
\begin{equation}
V_{\beta} (\textbf{q}) = \sum _j V_0
e^{ i \textbf{q} \cdot \textbf{r}_j}
\frac{2\sin (q_{y,x}R_j) \sin (q_{x,y}R_j)}{q_{y,x} \sin (2q_{x,y})},
\label{eq:Vb}
\end{equation}
and that for checkerboard modulations is
\begin{equation}
V_{\gamma} (\textbf{q}) = \sum _j V_0 e^{ i \textbf{q} \cdot \textbf{r}_j}
\left[ \frac{2\sin (q_y R_j) \sin (q_x R_j)}{q_y \sin (2q_x)}
+ \left( q_x \leftrightarrow q_y \right) \right]
\label{eq:Vc}
\end{equation}
Here all lengths are expressed in units of the lattice constant
$a$, $R_j$ is the averaged radius of the \textit{j}-th puddle, and
$V_0$ denotes the magnitude of the scattering potential by pinned
collective modes. For simplicity, we neglect the energy dependence
of $V_{\alpha,\beta,\gamma}$ and assume that $V_s$, $V_m$ and
$V_0$ are sufficiently small so that no resonance occurs in the
FT-LDOS.~\cite{ZhangD03} For sufficiently large scattering potentials,
full T-matrix calculations become necessary as in Ref.~\onlinecite{WangQH03}.
However, large $V_{s,m}$ would result in strong spectral asymmetry between
positive and negative bias voltages,~\cite{WangQH03} which differs from
experimental observation.~\cite{Hoffman02a,McElroy03,Howald03}
We also note that the energy dependence of
$V_{\beta,\gamma}$ reflects the spectral characteristics of the
collective modes and their interaction with quasiparticles and
impurities. For instance, we expect $V_{\gamma} \sim \zeta \gamma
^2$ for pinned SDW, where $\zeta$ is the impurity pinning strength
and $\gamma$ is the coupling amplitude of quasiparticles with SDW
fluctuations.~\cite{Polkovnikov02a,Polkovnikov03} Empirically for
nearly optimally doped Bi-2212, $R_j$ ranges from $5 \sim
10$.~\cite{Lang02} Here we take different values for $R_j$ with a
mean value $\langle R_j \rangle = 10$.

Given the Hamiltonian and the scattering potentials $V_{\alpha,
\beta, \gamma} (\textbf{q})$, we find that for infinite
quasiparticle lifetime and in the first-order T-matrix
approximation, the FT of the LDOS $\rho (\textbf{r},E)$ that
involves elastic scattering of quasiparticles from momentum
$\textbf{k}$ to $\textbf{k}+\textbf{q}$ is:
\begin{eqnarray}
&\rho _\textbf{q}(\omega) = - \frac{1}{\pi N^2} \lim _{\delta \to
0} \sum _\textbf{k} V_{\alpha, \beta
,\gamma} (\textbf{q}) \times \qquad \qquad \qquad \nonumber\\
 &\lbrace u_{\textbf{k}+\textbf{q}}u_\textbf{k}
\left( u_{\textbf{k}+\textbf{q}}u_\textbf{k}
\mp v_{\textbf{k}+\textbf{q}}v_\textbf{k} \right)
\Im \left[ \frac{1}{(\omega - E_\textbf{k} + i \delta )
(\omega - E_{\textbf{k}+\textbf{q}} + i \delta)} \right] \nonumber\\
 &+ u_{\textbf{k}+\textbf{q}}v_\textbf{k}
\left( u_{\textbf{k}+\textbf{q}}v_\textbf{k}
\pm v_{\textbf{k}+\textbf{q}}u_\textbf{k} \right)
\Im \left[ \frac{1}{(\omega + E_\textbf{k} + i \delta)
(\omega - E_{\textbf{k}+\textbf{q}} + i \delta )} \right] \nonumber\\
 &+ v_{\textbf{k}+\textbf{q}}u_\textbf{k}
\left( u_{\textbf{k}+\textbf{q}}v_\textbf{k}
\pm v_{\textbf{k}+\textbf{q}}u_\textbf{k} \right)
\Im \left[ \frac{1}{(\omega - E_\textbf{k} + i \delta)
(\omega + E_{\textbf{k}+\textbf{q}} + i \delta)} \right] \nonumber\\
 &- v_{\textbf{k}+\textbf{q}}v_\textbf{k}
\left( u_{\textbf{k}+\textbf{q}}u_\textbf{k}
\mp v_{\textbf{k}+\textbf{q}}v_\textbf{k} \right)
\Im \left[ \frac{1}{(\omega + E_\textbf{k} + i \delta)
(\omega + E_{\textbf{k}+\textbf{q}} + i \delta)} \right] \rbrace .
\label{eq:rhoq}
\end{eqnarray}
Here $N$ is the total number of unit cells in the sample, and
$\Im \lbrack \dots \rbrack$ denotes the imaginary part of the
quantity within the brackets, which is related to the
equal-energy quasiparticle joint density of states. The upper
(lower) sign in the coherence factor applies to spin-independent
(spin-dependent) interactions, $u_\textbf{k}$ and $v_\textbf{k}$
are the Bogoliubov quasiparticle coefficients,
$u_\textbf{k}^2 + v_\textbf{k} ^2 = 1$, $u_\textbf{k}^2 =
\left[ 1+ (\xi _\textbf{k}/E_\textbf{k}) \right]/2$,
$\xi _\textbf{k} \equiv \epsilon _\textbf{k} - \mu$,
$\epsilon _\textbf{k}$ is the tight-binding energy of the
normal state of Bi-2212 according to Norman \textsl{et al.}~\cite{Norman95},
$\epsilon _\textbf{k} = t_1 (\cos k_x + \cos k_y )/2 +
t_2 \cos k_x \cos k_y + t_3 (\cos 2 k_x +
\cos 2 k_y )/2 + t_4 (\cos 2 k_x \cos k_y +
\cos k_x \cos 2 k_y )/2 + t_5 \cos 2 k_x \cos 2 k_y$,
$t_{1-5} = -0.5951, 0.1636, -0.0519, -0.1117, 0.0510$ eV,
$\mu$ is the chemical potential, and $E_\textbf{k} =
\sqrt{\xi _\textbf{k} ^2 + \Delta _\textbf{k} ^2}$.

\begin{figure}[!thb]
\centering
\includegraphics[keepaspectratio=1,width=3.15in]{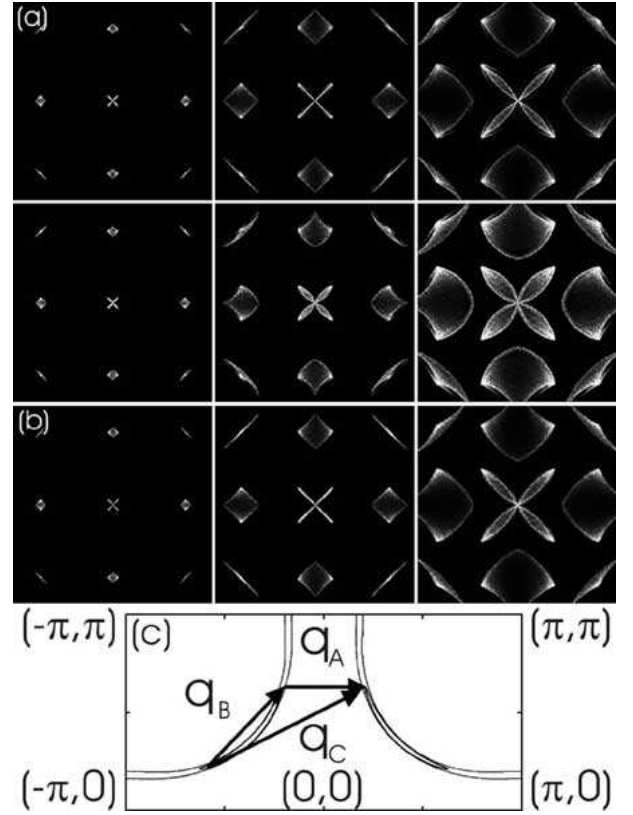}
\caption{Calculated energy-dependent Fourier transform (FT)
maps of quasiparticle LDOS in the first Brillouin zone with
randomly distributed non-magnetic point defects using Eq.~(\ref{eq:rhoq})
and $V_{\alpha}$: (a) $\Delta _d = 40$ meV and
$(\omega / \Delta _d) = \pm 0.15, \pm 0.45, \pm 0.75$ (up and down from
left to right); (b) $\Delta _d = 20$ meV
and $(\omega / \Delta _d) = 0.15, 0.45, 0.75$ (left to right).
(c) Schematic illustration of the equal-energy contours and
representative modulation wavevectors $\textbf{q}_A$,
$\textbf{q}_B$ and $\textbf{q}_C$, which correspond to $\textbf{q}_1$,
$\textbf{q}_7$ and $\textbf{q}_2$ in
Refs.~\onlinecite{Hoffman02a,McElroy03}.}\label{fig1}
\end{figure}

\begin{figure}[!thb]
\centering
\includegraphics[keepaspectratio=1,width=3.15in]{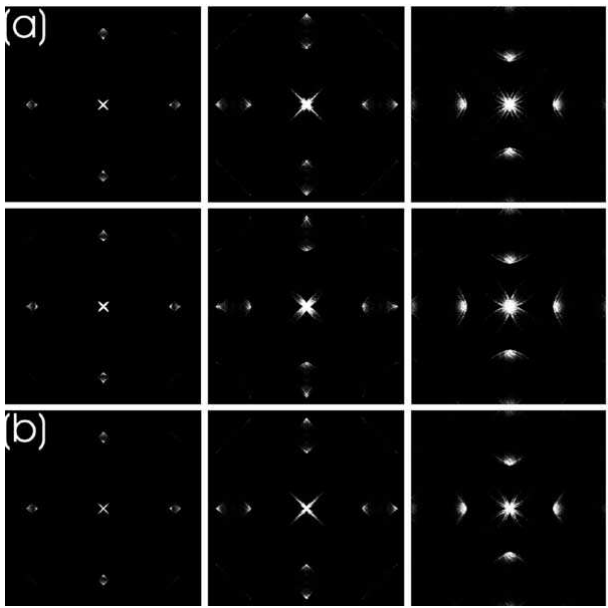}
\caption{Energy-dependent FT-LDOS maps with randomly distributed pinned
SDW using Eq.~(\ref{eq:rhoq}) and $V_{\gamma}$: (a) $\Delta _d = 40$ meV
and $(\omega / \Delta _d) = \pm 0.15, \pm 0.45, \pm 75$, up and down
from left to right; (b) $\Delta _d = 20$ meV and
$(\omega / \Delta _d) = 0.15, 0.45, 0.75$, from left to right.
The FT-LDOS does not exhibit discernible differences in the
spectral characteristics except the total intensities if we simply
replace $V_{\gamma}$ by $V_{\beta}$ and assume non-magnetic coherence
factors in Eq.~(\ref{eq:rhoq}).}
\label{fig2}
\end{figure}

\begin{figure}[!thb]
\centering
\includegraphics[keepaspectratio=1,width=3.15in]{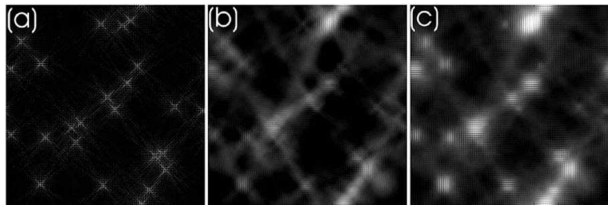}
\caption{Real space quasiparticle LDOS for a $(400 \times 400)$
area at $T = 0$ due to scattering by (a)non-magnetic point
impurities, (b) pinned CDW and (c) pinned SDW, for $\Delta _d =
40$ meV and $\omega = 30$ meV.} \label{fig3}
\end{figure}

\begin{figure}[!thb]
\centering
\includegraphics[keepaspectratio=1,width=3.15in]{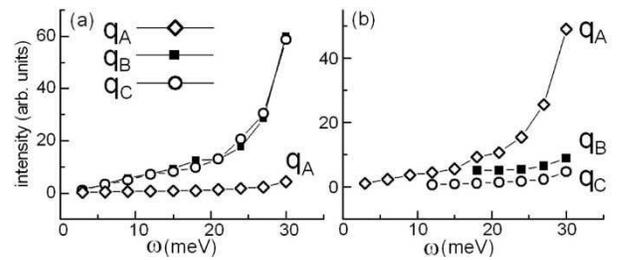}
\caption{Evolution of the relative intensities of FT-LDOS with
energy ($\omega$) for $q_A$, $q_B$ and $q_C$ as defined in
Fig.1(c) and $V_s$, $V_m$ and $V_0$ all taken to be unity:
quasiparticle scattering by (a) single non-magnetic point
impurity, and (b) single magnetic point impurity.} \label{fig4}
\end{figure}

Using Eq.~(\ref{eq:rhoq}) and $V_{\alpha,\beta,\gamma}
(\textbf{q})$, we obtain the energy-dependent FT-LDOS maps in the
first Brillouin zone for non-magnetic point impurities in
Fig.~\ref{fig1} with two different $\Delta _d$ values and for
pinned SDW (with spin-dependent coherence factor) in
Fig.~\ref{fig2}, whereas the corresponding LDOS modulations due to
$V_{\alpha,\beta,\gamma} (\textbf{q})$ in real space is shown in
Figs.~\ref{fig3}(a)-(c). For non-magnetic point impurity
scattering at $T \ll T_c$, the intensities associated with
$\textbf{q}_B$ and $\textbf{q}_C$ are much stronger than those of
$\textbf{q}_A$, as shown in Fig.~\ref{fig1} and also in
Fig.~\ref{fig4}(a). The results in Fig.~\ref{fig1} differ from the
STM observation~\cite{Hoffman02a,McElroy03} that reveals
comparable intensities associated with $\textbf{q}_A$ and
$\textbf{q}_B$, and weaker intensities with $\textbf{q}_C$.
Interestingly, the intensities of $\textbf{q}_A$ and
$\textbf{q}_{B,C}$ become reversed if one assumes magnetic point
impurity scattering, as illustrated in Fig.~\ref{fig4}(b).
However, there is no evidence of magnetic scattering in the
samples used in Refs.~\onlinecite{Hoffman02a,McElroy03}. In
contrast, the presence of pinned collective modes, regardless of
CDW or SDW, gives rise to much stronger intensities for
$\textbf{q}_A$ (by about two orders of magnitude), as shown in
Fig.~\ref{fig2}. Thus, the empirical FT-LDOS
maps~\cite{Hoffman02a,McElroy03} cannot be solely attributed to
quasiparticle scattering by non-magnetic point impurities.

\begin{figure}[!thb]
\centering
\includegraphics[keepaspectratio=1,width=3.15in]{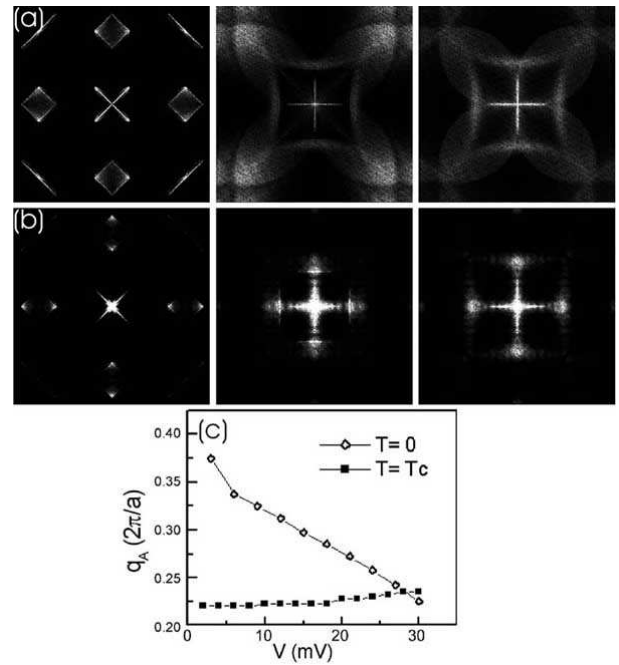}
\caption{The FT-LDOS maps at $T = 0$, $0.75 T_c$ and $T_c$ (from
left to right) for (a) point impurities $V_{\alpha}(q)$ and (b)
pinned SDW $V_{\gamma} (q)$. We assume $\Delta _d (T) = \Delta _d
(0) \lbrack 1 - (T/T_c) \rbrack ^{1/2}$, $\Delta _d (0) = 40$ meV,
tunneling biased voltage $=$ 18 mV, and $T_c = 80$ K. Besides
temperature dependent coherence factors, the thermal smearing of
quasiparticle tunneling conductance ($dI/dV$) is obtained by using
$(dI/dV) \propto |\int \rho _\textbf{q}(E) (df/dE)|_{(E-eV)} dE|$,
where $f(E)$ denotes the Fermi function. (c) $|q_A|$-vs.-$V$
(biased voltage) dispersion relation for pinned SDW at $T = 0$ and
$T_c$.} \label{fig5}
\end{figure}

The relevance of collective modes become indisputable when we
consider the temperature dependence of the FT-LDOS. As shown in
Fig.~\ref{fig5}(a), in the limit of $T \to T_c^-$, the $q$-values
contribute to the FT-LDOS map become significantly extended and
smeared for point-impurity scattering. In contrast, pinned SDW
yields strong intensities in the FT-LDOS map only at
$\textbf{q}_A$ for $T >\sim T_c$, as shown in Fig.~\ref{fig5}(b).
The overall energy dispersion due to pinned SDW is weaker than
that due to point impurities, as shown in Fig.~\ref{fig5}(c) for
$|\textbf{q}_A|$-vs.-$V$(biased voltage) at both $T = 0$ and $T =
T_c$. In particular, we note that the dispersion is further
reduced at $T_c$. These findings are consistent with recent
experimental observation by Yazdani \textsl{et
al}.~\cite{Yazdani03}

The energy, momentum and temperature dependence of our calculated
FT-LDOS in Figs.~\ref{fig1}-\ref{fig5} is supportive of spatially
modulated collective modes being relevant low energy excitations
in cuprates besides quasiparticles. In particular, only pinned
collective modes can account for the observation in the FT-LDOS
map above $T_c$. Although our simplified model cannot exclude CDW,
we note that pinned CDW would have coupled directly to the
quasiparticle spectra and resulted in stripe-like periodic local
conductance modulation, which has not been observed in STM
studies. On the other hand, various puzzling phenomena seem
reconcilable with the SDW scenario. For instance, the nano-scale
gap variations observed in Bi-2212~\cite{Lang02} may be understood
by noting that the LDOS in regions with disorder-pinned SDW
contains information of disorder potential coupled with
quasiparticles and SDW, so that the hump-like spectral
features at $\pm \Delta ^{\ast}$ represent neither the SDW gap nor
the superconducting gap $\Delta _d$, and the values of $\Delta
^{\ast}$ vary in accordance with the disorder potential. The
long-range spatial homogeneity of quasiparticle spectra in $\rm
YBa_2Cu_3O_{7-\delta}$ (YBCO)~\cite{Yeh01a} as opposed to the
strong spatial inhomogeneity in Bi-2212 can also be reconciled in
a similar context. That is, SDW can be much better pinned in
extreme two-dimensional cuprates like Bi-2212 than in more
three-dimensional cuprates such as YBCO. Furthermore, SDW can be
stabilized by magnetic fields,~\cite{Polkovnikov02a,Polkovnikov03,Demler01}
which naturally account for the checkerboard-like spectral modulations
around the vortex cores of Bi-2212.~\cite{Lake01,Hoffman02b}
Finally, the smooth evolution of
the pseudogap phase with temperature through $T_c$ may contribute
to the anomalously large Nernst effect under a c-axis magnetic
field above $T_c$,~\cite{WangY02} with spin fluctuations
responsible for the excess entropy.

In summary, we employ first-order T-matrix
approximation to study modulations in the quasiparticle
FT-LDOS of cuprates as a function of energy, momentum
and temperature. Our results suggest that a full account for
all aspects of experimental observation below $T_c$ must
include collective modes as relevant low energy excitations besides
quasiparticles, and that only collective modes can account
for the observed FT-LDOS above $T_c$.

\begin{acknowledgments}
We thank Professors Subir Sachdev, Doug Scalapino and C. S. Ting
and Mr. Yuan-Yu Jau for useful discussions. This research was
supported by NSF Grant \#DMR-0103045.
\end{acknowledgments}


\end{document}